\newcommand{\be}{\begin{equation}} \newcommand{\ee}{\end{equation}}
\newcommand{\bea}{\begin{eqnarray}}\newcommand{\eea}{\end{eqnarray}}
\newcommand{\nn}{\nonumber}
\newcommand\T{\mbox{Tr}}
\newcommand\s{\scriptscriptstyle}

\documentstyle[12pt]{article}
\begin{document}
\hfill hep-th/9610130 
\vspace{3cm}

\begin{center}
 {ZERO-CURVATURE REPRESENTATION AND DUAL FORMULATIONS OF $N=2$
 SUPERSYMMETRIC GAUGE THEORY IN HARMONIC SUPERSPACE}

\vspace{0.5cm}
{\large B.M.Zupnik}\\
{\it  Joint Institute for Nuclear Research,
 Dubna, Russia}
\vspace{0.5cm}

Talk at the XXI International Colloquium "Group Theoretical Methods in 
Physics", July 15-20, 1996, Goslar, Germany
\end{center}
\vspace{0.5cm}

\begin{abstract}
Analytical harmonic superfields are the basic variables of a standard
harmonic formalism of $SYM^2_4$-theory. We consider superfield actions for
 alternative formulations of this theory using the unconstrained harmonic
 prepotentials. The corresponding equations of motion are equivalent to the
 component field equations of $SYM^2_4$-theory. The analyticity condition 
appears only on-shell as the zero-curvature equation in the alternative
 formulations.
\end{abstract}

  Harmonic superspace has been introduced for the consistent off-shell
description of the supersymmetric theories with $N=2, D=4$ supersymmetry
\cite{GI1}. Analytical prepotentials of the $SYM^2_4$-theory $ V^{\s++}$
live in the analytic harmonic superspace with a restricted number of spinor
coordinates. The action of this theory is a nonlinear functional of the
analytic prepotentials \cite{GI2,Z1}.

We shall use the basic notions and notation of Ref\cite{GI2}.
Let us consider the harmonic-zero-curvature  equation for  the
harmonic connections $V^{\s++} , A^{\s--}$ with a dimension $d=0$
\be
\partial^{\s++}_{\s A} A^{\s--}
 - \partial^{\s--}_{\s A} V^{\s++} + [V^{\s++} , A^{\s--}] = 0
\label{A1}
\ee
This equation has the following perturbative solution \cite{Z1}:
\be
 A^{\s--}(V)=\sum\limits^{\infty}_{n=1} (-1)^n \int du_1
\ldots
du_n \frac{V^{\s++}(z,u_1 )\ldots V^{\s++}(z,u_n )}{(u^+ u^+_1)\ldots
 (u^+_n u^+ )}  \label{A2}
\ee
where the harmonic distributions $(u^+_1 u^+_2 )^{-1}$
 \cite{GI1} are used.

The nonlinear
equation of motion  of $SYM^2_4$ has the following form \cite{Z1}:
\be
F^{\s++}=  (D^+)^2(\bar{D}^+)^2 A^{\s--}(V) = 0
\label{A3}
\ee
where $D^+_\alpha$ and $\bar{D}^+_{\dot{\alpha}}$ are the flat spinor
derivatives in the harmonic superspace.

We use the analytic basis of the superfield gauge theory in the harmonic
superspace with the analytic gauge parameters $\Lambda$. This basis is
natural for the description of the covariantly analytic superfields.

Note that the analyticity conditions of the standard harmonic formalism
are kinematic (off-shell):
\be
D^+_\alpha V^{\s++}= \bar{D}^+_{\dot{\alpha}}V^{\s++}=0
\label{A4}
\ee

One should use also the conventional constraint  for the spinor
connection $A^-_a=(A^-_\alpha,\;\bar{A}^-_{\dot{\alpha}})$ \cite{Z1}
\be
[\nabla^{\s--},\;\nabla^+_a  ] =\nabla^-_a =D^-_a + A^-_a=
D^-_a - D^+_a A^{\s--}
\label{S4}
 \ee

Now we shall discuss the alternative (dual) formulation of the
$SYM^2_4$-theory . Let us consider the following representation
of the harmonic connection \cite{Z2}:
\be
A^{\s--}=\hat{A}^{\s--}=D^+_\alpha \;A^{\alpha\s (-3)} +
\bar{D}^+_{\dot{\alpha}}\;\bar{A}^{\dot{\alpha}\s(-3)}
=D^+_a \;A^{a\s (-3)} \label{A5}
\ee
where $A^{a\s(-3)}(z,u)$ are the unconstrained 4-spinor prepotentials
$a=(\alpha, \dot{\alpha})$ .
This representation solves explicitly Eq(\ref{A3}) but does not quarantees
the conservation of analyticity.

  The equation (\ref{A1}) in the new {\it $A$-frame} is treated as an
 integrable equation for the function  $V^{\s++}(A^{\s--})$ corresponding
to a choice of the independent variables (\ref{A5}) . Consider a dual in 
the $U(1)$-charge solution
\be
 \hat{V}^{\s++}(A^{a\s(-3)})=\sum\limits^{\infty}_{n=1} (-1)^n \int du_1
\ldots  du_n \frac{\hat{A}^{\s--}(z,u_1 )\ldots \hat{A}^{\s--}(z,u_n )}
{(u^- u^-_1)\ldots  (u^-_n u^- )}  \label{A6}
\ee
where the new harmonic distributions $(u^-_1 u^-_2 )^{-k}$ \cite{Z2} are
 introduced. These distributions satisfy the following relations:
\be
\partial^{\s--}_1 \frac{1}{(u^-_1 u^-_2 )^k} =\frac{1}{(k-1)!}
(\partial^{\s++}_1)^{k-1}\delta^{(-k,k)}(u_1 ,u_2 )
\label{A7}
\ee
which are completely analogous to the relations for the standard harmonic
distributions  with opposite $U(1)$-charges \cite{GI1}.

   Let us write the alternative superfield action of the $SYM^2_4$-theory
in the $A$-frame:
\be
S(A^{a\s(-3)}) =\int d^{12}z \sum\limits^{\infty}_{n=2} \frac{(-1)^n}{n}
 \int du_1 \ldots  du_n \frac{\T[\hat{A}^{\s--}(z,u_1 )\ldots 
\hat{A}^{\s--}(z,u_n )]}  {(u^- u^-_1)\ldots  (u^-_n u^- )}  \label{A8}
\ee
where $\hat{A}^{\s--}= D^+_a \;A^{a\s (-3)} $.

Consider an arbitrary variation of this functional
\be
\delta S =\int d^{12}z du\T \left[\delta\hat{A}^{\s--}\;
 \hat{V}^{\s++}\right]=\int d^{12}z du\T \left[\delta A^{a\s(-3)}\;
D^+_a \hat{V}^{\s++}\right] \label{A9}
\ee

Thus, the action (\ref{A8}) generates a dynamical analyticity equation
of $SYM^2_4$ \cite{Z2}:
\be
D^+_a \hat{V}^{\s++}(A^{a\s(-3)})=0  \label{A10}
\ee

Thus, the nonlinear dynamical equation (\ref{A3}) and the kinematic
analyticity of the standard harmonic formalism correspond
 to the linear constraint on the dual harmonic-superfield variable
$A^{\s--}$ and the dynamical zero-curvature equation.
Note, that the chirality properties and the full set of Bianchi identities
 for the $SYM^2_4$ tensors $W=(\bar{D}^+)^2 A^{\s--}$ and
 $\overline{W}=(D^+)^2 A^{\s--}$ are the consequence of the dynamical
 analyticity and arise only on-shell in the dual formulation.

The $A$-frame prepotential possesses the following gauge transformations
\cite{Z2}:
\be
\delta A^{a\s(-3)}= R^{a\s(-3)}\Lambda + [\Lambda, A^{a\s(-3)}] + D^+_b
\Lambda^{ab\s(-4)} \label{A11}
\ee
where a general symmetrical spinor $\Lambda^{ab\s(-4)}$ and an analytic
scalar $\Lambda$ are the Lie-algebra valued superfield gauge parameters
and $R^{a\s(-3)}$ is some spinor differential operator.
The spinor derivative of $\delta A^{a\s(-3)}$ produces the standard
gauge transformation of the harmonic connection
\be
\delta \hat{A}^{\s--} = \partial^{\s--}\Lambda +\left[\hat{A}^{\s--},
\Lambda\right] =\nabla^{\s--}\Lambda
  \label{A12}
\ee

The $A$-frame action is invariant under the  gauge transformations
of new prepotentials
\bea
&\delta_\Lambda S =\int d^{12}z du\T [\nabla^{\s--}\Lambda
 \hat{V}^{\s++}]=&\\ \nn
&-\int d^{12}z du\T \left[\Lambda \nabla^{\s--}\hat{V}^{\s++}\right]=
-\int d^{12}z du\T \left[\Lambda \partial^{\s++} D^+_a \;A^{a\s (-3)}
\right]=0&
 \label{A13}
\eea

It should be stressed that prepotential $A^{a\s(-3)}$ of our harmonic
formalism contains an infinite number of harmonic
 auxiliary fields , the physical component $SYM^2_4$ fields and pure
gauge degrees of freedom in contrast with the analytic prepotential of
the standard harmonic approach which has physical and pure gauge 
components only \cite{GI1}. The physical sector of the superfield 
$D^+_a \;A^{a\s (-3)}$
 contains the vector $A_{\alpha\dot{\beta}}$, the spinor $\psi^a_i $,
  the independent field-strengths $F^{\alpha\beta},\;
\bar{F}^{\dot{\alpha}\dot{\beta}}\;$, $f^{\alpha\dot{\beta}},\;
\bar{f}^{\dot{\alpha}\dot{\beta}}$ and the scalars $\Phi,\;\bar{\Phi}$.
The dynamical analyticity condition (\ref{A10}) is equivalent to
the component $SYM^2_4$  equations of motion.
 All  auxiliary fields vanish on-shell and
 the standard first-order component $SYM^2_4$ equations for
 $F,\bar{F},f,\bar{f},  A$ and $\psi$ arise, too. It is evident that the
different representations  of
 $SYM^2_4$ theory are equivalent on-shell and have identical component
 solutions for the physical fields.

One can consider the intermediate version of the harmonic-superfield
action of $SYM^2_4$ which interpolate between the action with the analytic
prepotential and the dual action (\ref{A8}). Introduce the  independent
nonanalytic harmonic connection $\widetilde{V}^{\s++}$
\be
\delta\widetilde{V}^{\s++}=\partial^{\s++}\Lambda +\left[\widetilde{V}^
{\s++},\Lambda\right] \label{A14}
\ee

The intermediate version of the $SYM^2_4$-action has the following form:
\bea
 S(\widetilde{V}^{\s++},A^{a\s(-3)}) =\sum\limits^{\infty}_{n=2} 
\frac{(-1)^n}{n}\int d^{12}z du_1 \ldots du_n
\frac{\T \left[\widetilde{V}^{\s++}(z,u_1)\ldots \widetilde{V}^{\s++}
(z,u_n)\right]}{(u_1^+ u_2^+) \ldots (u_n^+u_1^+)}-\; &\\ \nn
-\int d^{12}z du \T \left[D^+_a \;A^{a\s (-3)} \widetilde{V}^{\s++}\right]
\hspace{5cm}&\label{A15}
 \eea
where the dual superfield $A^{a\s (-3)}$ is treated as a Lagrange 
multiplier.
This action is invariant under the analytical gauge transformations of
$\widetilde{V}^{\s++}$ and $\hat{A}^{\s--}$ with the common analytical 
parameters.

The corresponding equations of motion are
\be
\widetilde{A}^{\s--}(\widetilde{V}^{\s++})=D^+_a \;A^{a\s (-3)}
\label{A16}
\ee
where one should use the analogue of Eq(\ref{A2}) in the left-hand side and
\be
D^+_a \widetilde{V}^{\s++} =0  \label{A17}
\ee

One can discuss also the 2nd alternative form of the interpolating
 $SYM^2_4$-action. Let an unconstrained harmonic superfield 
${\cal A}^{\s--}(z,u)$ 
 and the analytic harmonic superfield $V^{\s++}$  are the independent
 connections  with the standard transformation laws. We suppose
that Eq(\ref{A1}) is not valid off-shell in this formulation.
The new interpolating action has the following form:
\bea
 S({\cal A}^{\s--},V^{\s++}) =\sum\limits^{\infty}_{n=2}\frac{(-1)^n}{n} 
  \int d^{12}z du_1 \ldots du_n
\frac{\T \left[{\cal A}^{\s--}(z,u_1)\ldots {\cal A}^{\s--}(z,u_n)
\right]}{(u_1^- u_2^-) \ldots (u_n^-u_1^-)}-\; &\\ \nn
-\int d^{12}z du \T \left[{\cal A}^{\s--} V^{\s++}\right]
\hspace{5cm}&\label{A18}
 \eea
where $(u_1^- u_2^-)^{-1}$ is the  harmonic distribution (\ref{A7}).

      The corresponding harmonic-superfield equations of motion  are
 equivalent to Eqs(\ref{A1},\ref{A3}) of the standard harmonic formalism.
 It is easy to obtain the superfield Feynman rules for this  action
  by the analogy with Ref\cite{GI2}.

We hope that the dual formulations of   $SYM^2_4$ in the harmonic
superspace will be useful for the analysis of the remarkable quantum
properties of this theory.

 The author is grateful to A.S.Galperin and E.A.Ivanov for stimulating 
discussions. This work is partially supported by INTAS-grant 94-2317 and
 the grant of the Dutch NWO organization.

\end{document}